\documentclass[14pt]{article}
\usepackage{hyperref}
\usepackage{arxiv}
\usepackage{ulem}
\usepackage{xspace}
\usepackage{graphicx}
\usepackage{fancyhdr}
\usepackage{pslatex}
\usepackage{xcolor}
\usepackage{overpic}
\usepackage{bm}
\usepackage{epsfig}
\usepackage{pdfpages}
\usepackage{graphics}
\usepackage{color}
\usepackage{amsmath, amsthm, amssymb}
\usepackage{amssymb}
\usepackage{enumerate}
\usepackage{enumitem}
\usepackage[symbol]{footmisc}
\usepackage{dsfont}
\usepackage{cite}

\usepackage[utf8]{inputenc} % allow utf-8 input
\usepackage[T1]{fontenc}    % use 8-bit T1 fonts
\usepackage{epstopdf}

\usepackage{caption}
\title{Test of local realism via entangled $\Lambda\bar\Lambda$ system\footnotemark}
\author{
M.~Ablikim$^{1}$, M.~N.~Achasov$^{5,b}$, P.~Adlarson$^{74}$, X.~C.~Ai$^{80}$, R.~Aliberti$^{35}$, A.~Amoroso$^{73A,73C}$, M.~R.~An$^{39}$, Q.~An$^{70,57}$, Y.~Bai$^{56}$, O.~Bakina$^{36}$
% ...
\\
\vspace{0.2cm}
(BESIII Collaboration)\\
\vspace{0.2cm}{\it
$^{1}$ Institute of High Energy Physics, Beijing 100049, People's Republic of China. \\
% ...
$^{5}$ Budker Institute of Nuclear Physics SB RAS (BINP), Novosibirsk 630090, Russia. \\
% ...
$^{35}$ Johannes Gutenberg University of Mainz, Johann-Joachim-Becher-Weg 45, D-55099 Mainz, Germany. \\
$^{36}$ Joint Institute for Nuclear Research, 141980 Dubna, Moscow region, Russia. \\
% ...
$^{39}$ Liaoning Normal University, Dalian 116029, People's Republic of China.\\
% ...
$^{56}$ Southeast University, Nanjing 211100, People's Republic of China. \\
   $^{57}$ State Key Laboratory of Particle Detection and Electronics, Beijing 100049, Hefei 230026, People's Republic of China.\\
% ...
$^{70}$ University of Science and Technology of China, Hefei 230026, People's Republic of China. \\
% ...
$^{73}$ University of Turin and INFN, (A)University of Turin, I-10125, Turin, Italy; (B)University of Eastern Piedmont, I-15121, Alessandria, Italy; (C)INFN, I-10125, Turin, Italy. \\
     $^{74}$ Uppsala University, Box 516, SE-75120 Uppsala, Sweden. \\
% ...
$^{80}$ Zhengzhou University, Zhengzhou 450001, People's Republic of China.\\
\vspace{0.2cm}
% ...
$^{b}$ Also at the Novosibirsk State University, Novosibirsk, 630090, Russia. \\
% ...
}
}

\makeatletter
\renewcommand{\maketitle}{
  \begin{center}
    {\LARGE\@title\par}
    \vskip 1em
    {\large\@author\par}
  \end{center}
  \par\vskip 1.5em
}
\makeatother
\makeatletter
\renewcommand{\@makefnmark}{\hbox{\textsuperscript{*}}}  % 标记显示 *
\makeatother
\begin{document}
\maketitle
\footnotetext{This paper will appear in Nature Communications.}
\begin{abstract}
The non-locality of quantum correlations is a fundamental feature of
quantum theory. The Bell inequality serves as a benchmark for distinguishing between predictions made by quantum theory and local hidden variable theory (LHVT). Recent advancements in photon-entanglement experiments have addressed potential loopholes and
have observed significant violations of variants of Bell inequality. However, examples of Bell inequalities violation in high energy physics are scarce. In this study, we utilize $(10.087\pm0.044)\times10^{9}$ $J/\psi$ events collected with the BES-III detector at the BEPCII collider, performing non-local correlation tests using the entangled hyperon pairs. The massive-entangled $\Lambda\bar\Lambda$ systems are formed and decay through strong and weak interactions, respectively. Through measurements of the angular distribution of $p\bar{p}$ in $J/\psi\to \gamma\eta_c$ and subsequent $\eta_c\to\Lambda(p\pi^-)\bar\Lambda(\bar{p}\pi^{+})$ cascade decays, a significant violation of LHVT predictions is observed. The exclusion of LHVT is found to be statistically significant at a level exceeding $5.2\sigma$ in the testing of three Bell-like inequalities.
\end{abstract}

\section{Introduction}
Of all the fundamental aspects of quantum mechanics (QM), perhaps the most bizarre is the non-local nature of an entangled system consisting of two or more components, whose quantum state cannot be factored into
the tensor product of the quantum state of each individual
member. ``Local'' and ``locality'' mean that an object is influenced
only by its surroundings and that any influence cannot travel faster
than the speed of light.  Consider two observers, Alice and Bob, who
are spatially far apart from each other and possess half of an
entangled quanta pair. According to QM, a measurement by Alice can
instantaneously affect the state of her partner, and vice versa. This
entanglement between observers is non-local, superluminal, and
seemingly incompatible with Special Relativity.  Einstein called this
``spooky action at a distance''.  In 1935, Einstein, Podolsky, and
Rosen devised a thought experiment, known as the EPR
paradox~\cite{Einstein:1935rr}.  The paradox led EPR to conclude that
"the description of reality given by the wave function in QM is not
complete''~\cite{Einstein:1935rr}, suggesting the existence of an local
hidden variable theory (LHVT).

In 1951, Bohm modified the EPR paradox to make it experimentally
accessible~\cite{Bohm:1951,Bohm:1957zz}. In Bohm's scheme, a pair of
entangled spin-1/2 particles in a spin singlet state are used. If
Alice measures the spin of her particle to be along the $z$ direction,
Bob should find the spin of his along the $-z$ direction.
In 1964, Bell developed his own variant of the EPR
paradox~\cite{Bell:1964}. Bell assumed that the measurement results of
Alice and Bob could be described by two families of variables. The
outcomes of their measurements in space-like separation do not affect
each other and are mutually independent (local). Under these
assumptions, he provided an upper bound for hidden-variable
correlation according to the Bell inequality, which QM can
violate in specific regions of parameter space. Bell showed that QM is incompatible with LHVT, which was known as the Bell
Theorem. The triumph of the Bell inequality lies in transforming the philosophical debates on the
completeness of QM into an experimental criterion.

Numerous optical experiments using entangled photons have been
conducted to test Bell inequalities
\cite{Freedman:1972,Aspect:1981,Aspect:1982,Weihs:1988,Rowe:2001,Pan:2000,Marissa:2015}.
However, most experiments relied on additional assumptions in order to exclude LHVTs, therefore not closing all the so-called loopholes. There are three commonly admitted loopholes
\cite{lpstyle}. The locality loophole means the separation of Alice
and Bob is not space-like. By increasing the distance between them and
shortening the interval of successive measurements, the space-like separation requirement can be met, ensuring no physical information is
exchanged between Alice and Bob, even by light. The freedom-of-choice
loophole addresses whether Alice and Bob can freely and independently
decide what to measure. LHVT postulates that measurements can be
performed with mutual independence~\cite{Bell:2004}. The third loophole, known as the fair sampling loophole (or detection loophole)~\cite{Pearle:1970,Brunner:2014,Larsson:2014}, can occur when a subset of detected particles violates a Bell inequality while the total group does not. If an experiment only detects this subset and assumes it represents the entire particels, a loophole is exploited. Closing this loophole is possible by detecting the particles with sufficient efficiency, which was realized in optical experiments\cite{Marissa:2015, Stevens:2015awv}.
Optical experiments designed to test Bell inequalities have made significant
progress in closing potential loopholes
\cite{Brunner:2014,Larsson:2014} since the pioneering work~\cite{nobelprize}. These
experiments, both with specific loopholes and without, have produced
results that clearly violate the Bell inequalities
\cite{Freedman:1972,Aspect:1981,Aspect:1982,Weihs:1988,Rowe:2001,Pan:2000,Marissa:2015}.

Unlike experiments with entangled photons, studies utilizing entangled-massive particles to investigate local realism~\cite{Rowe:2001,Sakai:2006,Arunav:2022,Colciaghi:2023} are uncommon, and these experiments in low energy region often do not address all three loopholes simultaneously. The detection loophole was addressed in the entangled ion experiments \cite{Rowe:2001,Sakai:2006} with significant results. The entangled states are prepared with specific ion traps, or nuclear reactions. Here, we present an experiment testing realism with entangled $\Lambda\bar\Lambda$ particles. It has three eminent features: the first is that the entangled $\Lambda\bar\Lambda$ particles are realized over extremely short distances, accompanied by strong and weak interactions in the decays, severing as a spin self-analyzer. The second feature is that the $\Lambda\bar\Lambda$ particles is the maximally entangled states \cite{Wootters:1997id} produced from the $\eta_c$ decays. The third is concerning the locality loophole, which can be addressed by requiring the space-like criteria applied to the decayed $\Lambda\bar\Lambda$ particles. Our experiment does not close the detection loophole, since having a fair sampling of the recorded events is assumed. However, it is a widely accepted fundamental assumption in collider physics \cite{Fabbrichesi:2024rec}.

Testing LHVT in high energy physics are challenging, and has garnered substantial attention
over the past two decades \cite{Li:2010, Qian:2020, Privitera:1991nz, Hao:2010,Chen:2012,Li:2006fy,Hiesmayr:2011na,Baranov:2008,Chen:2013epa, Abel:1992kz,Go:2004,Genovese:2001ze,Barr:2024djo,Fabbrichesi:2024prd}. The
proposed experiments can be categorized into two groups, quark flavor
entanglement (also known as quasi-spin) experiments
\cite{Hiesmayr:2011na,Go:2004,Genovese:2001ze} and particle spin
entanglement experiments
\cite{Qian:2020,Privitera:1991nz,Hao:2010,Chen:2012,Li:2006fy,Baranov:2008,Chen:2013epa,Abel:1992kz,Horodecki:1995nsk,Fabbrichesi:2024prd}. Tornquist
proposed to examine the
non-locality of quantum mechanical prediction using the spin-entangled
$\Lambda\bar\Lambda$ system~\cite{Tornqvist:1980af}. The decay
$\Lambda(\bar\Lambda)\to p\pi^-(\bar p\pi^+)$ can serve as a spin
analyzer for inferring the hyperon spins
\cite{Hiesmayr:2014jva}. Specifically, the spins of $\Lambda$ and
$\bar\Lambda$ in the $\eta_c\to\Lambda\bar\Lambda$ process are always
opposite and possess a total spin of 0, i.e., $|S\rangle =
1/\sqrt{2}( |\uparrow \downarrow\rangle-|\downarrow \uparrow\rangle)$,
where $\uparrow$ and $\downarrow$ denote the spin projections of
$\Lambda(\bar\Lambda)$.  Due to parity violation in hyperon weak
decays, the outgoing proton (anti-proton) exhibits a preference to
travel along (against) the polarization direction of the hyperon.
Alice and Bob arrange to measure the spins of their respective particles at the
$\Lambda$ and $\bar{\Lambda}$ decay vertices, with their measurement axes setting aligned
along the momentum directions of the proton and antiproton. The correlation
function between these measurements can be expressed as~\cite{Tornqvist:1980af}:
\begin{equation}
E(\vec{n}_1, \vec{n}_2) \equiv \langle S|\sigma\cdot \vec{n}_1\sigma\cdot \vec{n}_2|S\rangle= -\vec{n}_1 \cdot \vec{n}_2 \equiv -\cos\theta_{p\bar{p}},
\end{equation}
 where the operator $\sigma\cdot \vec{n}_1(\sigma\cdot \vec{n}_2)$ represents the measurement of
$\Lambda(\bar\Lambda)$ spin projection along the guide axis
$n_1(n_2)$, which coincides with the proton (antiproton) momentum direction in the
$\Lambda$ ($\bar{\Lambda}$) rest frame. This direction is obtained by boosting the proton(antiproton) momentum to the $\Lambda(\bar\Lambda)$ center-of-mass system. Here $\theta_{p\bar{p}}$ is the opening angle between $\vec{n}_1$ and $\vec{n}_2$ with their reference frames superimposed.
 Thus, LHVT can be experimentally tested by measuring the
distribution $I(\theta_{p\bar{p}})$ of angles between the momenta of
$p$ and $\bar p$~\cite{Tornqvist:1980af},
\begin{equation}
I(\theta_{p\bar{p}}) =1+\alpha^2\cos\theta_{p\bar{p}},
\end{equation}
where $\alpha=\alpha_\Lambda= 0.750 \pm 0.009 \pm 0.004$ is the
asymmetry parameter of the $\Lambda\to p\pi^-$ decay, which can be precisely
measured in $J/\psi\to\Lambda(p\pi^-)\bar\Lambda(\bar p\pi^+)$
~\cite{BESIII:2018cnd,BESIII:2022prl}. If a hidden measurement of $\Lambda$ polarization is carried out before its decay, this reduces to $\alpha^2=\alpha_\Lambda^2/3$. When the Bell inequality is applied to
the decay $\eta_c\to\Lambda\bar\Lambda$, one may get a bound \cite{Tornqvist:1980af} as
$|E(\vec{n}_1,\vec{n}_2)|\le 1-{2\theta_{p\bar{p}}\over \pi}$. Substituting with
$I(\theta_{p\bar p})$, we obtain:
\begin{equation}\label{EQ::Tornquist}
|I(\theta_{p\bar{p}}) -1|\le \alpha_\Lambda^2(1-{2\over \pi}\theta_{p\bar{p}}),
\end{equation}
which defines the domain satisfying the Bell inequality. If experimental measurement of the angular distribution $\cos\theta_{p\bar p}$ lies outside the region allowed by Bell’s inequality, then a violation of Bell inequality is established.

Later, a freedom-of-choice loophole was identified when testing the Bell inequality
using $\Lambda\bar\Lambda$ spin entanglement produced in $\eta_c$
decays~\cite{Hiesmayr:2014jva,Selleri:1988tf}. Compared to optical
experiments, the decays of $\Lambda$ and $\bar\Lambda$ occur
spontaneously, rather being artificially controlled by
experimenters at will. The assumption of independence in the decay of $\Lambda$ and $\bar\Lambda$ particles is utilized during realism testing. Closing this loophole in high-energy experiments is challenging, as suggested in Ref. \cite{Hiesmayr:2011na}, requiring dedicated devices for active measurements.
Although Ref.~\cite{Fabbrichesi:2024rec} suggests that measurements of $\Lambda$ and $\bar\Lambda$ decays in the detector could serve as an ideal random generator, the freedom-of-choice loophole remains a significant challenge in high-energy experiments.
Nevertheless, we could introduce a weak assumption that the sample survived from the free-will choice should have the same distribution as the detected sample. Thus the presence or absence of free-will choice only affects the sensitivity to test realism, but does not alter the acceptance or rejection of realism conclusions in the high luminosity of collider experiment.

It should be noted that in previous scheme the $\Lambda(\bar\Lambda)$ decay parameter $\alpha_\Lambda(\alpha_{\bar\Lambda})$ is introduced in the test of Bell inequality, which will bring about a new loophole. In order to overcome the QM dependence, a new inequality, a Clauser-Horne (CH)-type inequality, was developed~\cite{Qian:2020,Shi:2019kjf}. For the
$\eta_c\to\Lambda\bar\Lambda$ decay, the CH inequality can be
generalized as~\cite{Qian:2020}
\begin{equation}
P(\vec n_a,\vec n'_b)-P(\vec n_a,\vec n'_d)+P(\vec n_c, \vec n'_b)+P(\vec n_c,\vec n'_d)-P(\vec n_c)-P(\vec n'_b)+{1-\alpha_\Lambda^2\over 2}\le (1-\beta_\Lambda){\alpha^2_\Lambda\over 2}.
\end{equation}
Here unit vectors $\vec n_{a,c}$ and $\vec n'_{b,d}$ denote the directions of chosen guide axes used to detect proton and antiproton, respectively. This can be viewed as a generalization of polarizer settings in optical experiments. $P(\vec n, \vec n')={1\over 4}(1+\alpha_\Lambda^2 \vec n\cdot\vec n')$ represents
the probability to detect an event of proton and antiproton at direction $\vec n$ and $\vec n'$ respectively, coinciding with a $\eta_c\to \Lambda(p\pi^-)\bar\Lambda(\bar p\pi^+)$ decay. While $P(\vec n_c)=P(\vec n'_b)={1\over 2}$ indicates the probability to detect proton or antiproton alone. Just as the CH inequality is tested in optical
experiments, these direction settings are used to characterize the
polarization of $\Lambda(\bar\Lambda)$.  In the above inequality,
$\beta_\Lambda = P/E$, is the velocity of $\Lambda$, and $P$ and
$E$ are the $\Lambda$ momentum and energy, respectively. The introduction of $\beta_\Lambda$ is necessary to account for the requirement of space-like separation, which decreases the upper bound. The nonzero upper bound of
the CH inequality is due to the requirement, described below, of
excluding any possible classical communication between the $\Lambda$
and $\bar\Lambda$. To directly verify the contradiction between
locality and QM, one can obtain the CH inequality by
substituting the QM predictions into the above
equation. Specifically, the CH inequality is given by:
\begin{equation}
{\alpha_\Lambda^2\over 4}(\cos\theta_{ab}-\cos\theta_{ad}+\cos\theta_{cb}+\cos\theta_{cd})-{\alpha_\Lambda^2\over 2}\le  (1-\beta_\Lambda){\alpha^2_\Lambda\over 2},
\end{equation}
where $\theta_{ij}$ are the angles between $\vec n_i$ and $\vec n'_j$.
To highlight the specific region where the violation of the CH inequality occurs, one selects $\theta_{ab}=\theta_{cb}=\theta_{cd}=\theta_{p\bar p}$ and $\theta_{ad}=3\theta_{p\bar p}$, resulting in the following expression~\cite{Qian:2020}:
\begin{equation}\label{newchi}
CH(\theta_{p\bar{p}})\equiv\alpha_\Lambda^2\left[ {3\cos\theta_{p\bar{p}} -\cos(3\theta_{p\bar{p}})\over 4}-{1\over 2}\right]\le  (1-\beta_\Lambda){\alpha^2_\Lambda\over 2} .
\end{equation}
Since $\alpha_\Lambda^2>0$, the above inequality may test the LHVT evidently independent of $\alpha_\Lambda$, superior to inequality (\ref{EQ::Tornquist}). From the
generalized CH inequality, the maximum violation of the inequality,
$\alpha_\Lambda^2({\sqrt2\over 2}-{1\over 2})\le
(1-\beta_\Lambda){\alpha^2_\Lambda\over 2}$ with $\beta_\Lambda\sim
0.7$, is achieved when $\theta_{p\bar{p}} = \pi/4$. If a significant number of events can be observed
near $\theta_{p\bar{p}} = \pi/4$ with $CH(\theta_{p\bar
  p})>(1-\beta_\Lambda){\alpha^2_\Lambda\over 2}$, the prediction of
quantum theory is inconsistent with the locality.
\section{BESIII EXPERIMENT AND MONTE CARLO SIMULATION}
The BESIII detector~\cite{besiii} records symmetric $e^+e^-$
collisions provided by the BEPCII storage ring~\cite{lum}, which
operates with a peak luminosity of 1$\times$10$^{33}$
cm$^{-2}$s$^{-1}$ in the center-of-mass energy range from 2.0 GeV to
4.95 GeV. The cylindrical core of the BESIII detector covers 93\% of
the full solid angle and consists of a helium-based multilayer drift
chamber (MDC), a plastic scintillator time-of-flight system (TOF), and
a CsI(Tl) electromagnetic calorimeter (EMC), which are all enclosed in
a superconducting solenoidal magnet providing a 1.0 T (0.9 T for 2012 $J/\psi$ data) magnetic field. The tracks of charged particles are reconstructed and their momenta are determined in the MDC, while showers from photons are
reconstructed and their energy deposits are measured in the EMC.

Monte Carlo (MC) simulations are used to optimize the event selection criteria and
estimate the background sources, as well as to determine the
efficiency. {\sc geant4}~\cite{geant4} based MC software, including
the geometric description of the BESIII detector~\cite{descrBESIII,descrdisplay}
and its response, is used to simulate the MC samples. The inclusive MC
sample includes the production of vector charmonium(-like) states and
the continuum processes incorporated in {\sc kkmc}
\cite{Jadach:1999vf}. All particle decays are modelled with {\sc
  evtgen}~\cite{Lange:2001uf,Ping:2008zz} using the branching
fractions either taken from the Particle Data Group
\cite{ParticleDataGroup:2022pth}, or otherwise estimated with {\sc
  lundcharm}~\cite{lund1, lund2}.

\section{EVENTS SELECTION}
In this work, an experimental test of the non-local
correlation in the $\Lambda\bar\Lambda$ system is performed using
$(10.087\pm0.044)\times10^{9}$ $J/\psi$ events collected by the BESIII
detector at the BEPCII $e^+e^-$ collider~\cite{Njpsi:2022} through
$J/\psi\to\gamma\eta_c\to\gamma\Lambda(p\pi^{-})\bar\Lambda(\bar{p}\pi^+)$
decays, and the event selection criteria are described in the Appendix below. The locality loophole is closed by applying a requirement on
the hyperon decay length to guarantee the spatial separation between
their decays. From the $e^+e^-$ interaction point, where $\eta_c$
decays into $\Lambda$ and $\bar\Lambda$ instantaneously, the average
flight distance of $\Lambda$ ($\bar\Lambda$) to the decay point is
about 6.95 cm. The
$\Lambda$ and $\bar\Lambda$ candidates are identified by fitting the
secondary vertices to $p\pi^-$ and $\bar{p}\pi^+$ final states, respectively. The fits yield
the decay lengths $L_1$ and $L_2$, which are the flight distances of the
$\Lambda$ and $\bar\Lambda$ from the beam interaction point,
respectively. Testing LHVT requires space-like separation of $\Lambda$
and $\bar\Lambda$, and $L_1$ and $L_2$ are used to select separated
events~\cite{Tornqvist:1980af,Qian:2020}, i.e.
\begin{equation}\label{eq:space}
{1\over S}\le {L_1\over L_2}\le S, \text{~with~} S={1+\beta_\Lambda\over 1-\beta_\Lambda}.
\end{equation}
After applying the above selection criteria, 23,313 events survived. The detection efficiency was determined to be 8.2\%. The number of background events was estimated to be 4,319, mainly from the decays of $J/\psi\to \bar\Lambda\Sigma^0(\gamma\Lambda)+c.c.,~\gamma\bar\Lambda\Sigma^0(\gamma\Lambda)+c.c.,~\Sigma^0(\gamma\Lambda)\bar\Sigma^0(\gamma\bar\Lambda),\text{and}~\pi^0(2\gamma)\Lambda\bar\Lambda$.

\begin{figure}[htbp]
\vspace{1cm}
\begin{center}
\begin{overpic}[width=0.6\textwidth]{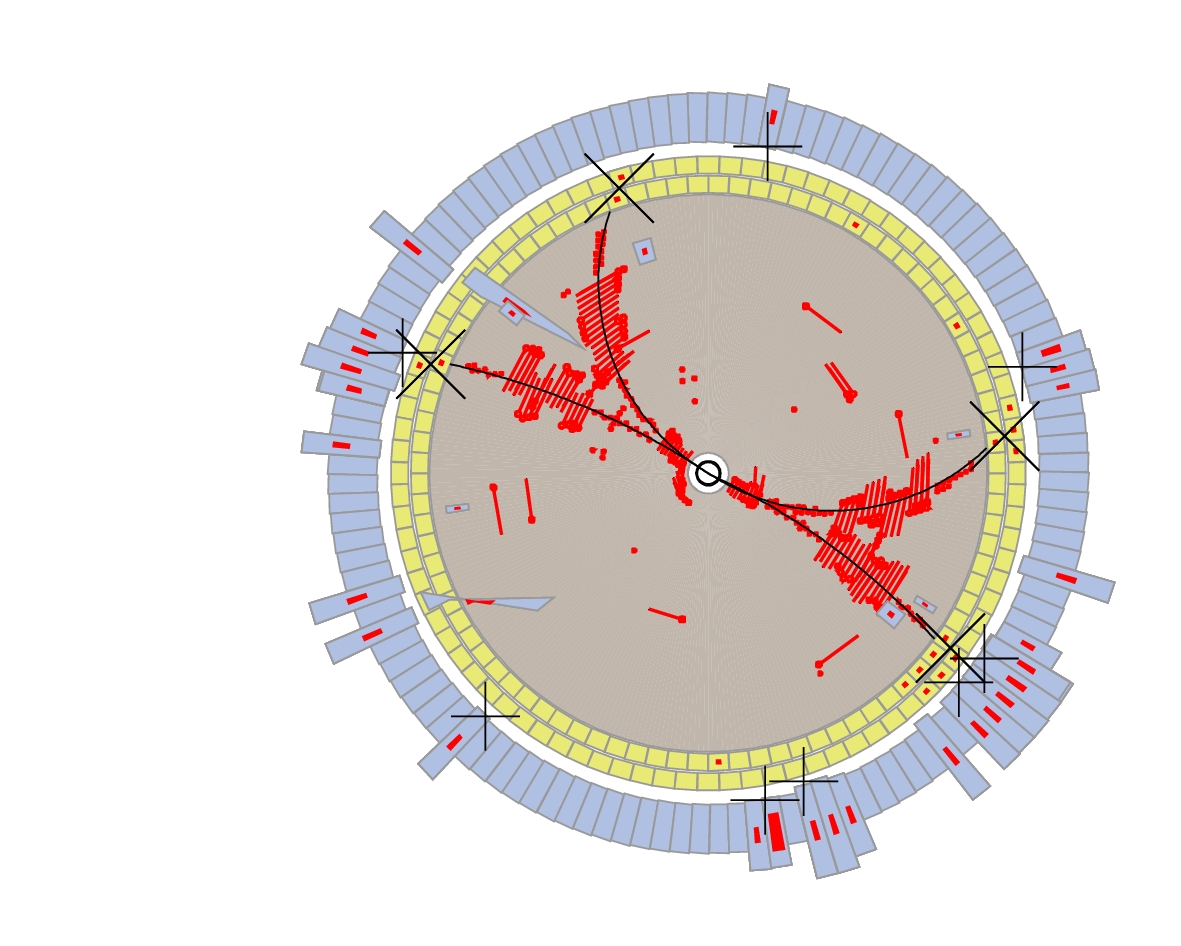}
\end{overpic}
\caption{{\bf| Transverse view of a $\Lambda\bar\Lambda$ space-like
    separation event in the detector.} The center of the detector
  corresponds to the $e^+e^-$ interaction point, where the $J/\psi$ is
  produced and decays instantaneously into a photon and the $\eta_c$
  meson. The $\eta_c$ itself decays also instantaneously into a
  hyperon pair $\Lambda\bar{\Lambda}$, and the $\Lambda(\bar{\Lambda})$ decays
  into the charged particles $p$ and $\pi^-$ ($\bar{p}$ and
  $\pi^+$). The trajectories of the charged tracks can
  be seen as black curves. The areas colored
  beige, yellow and light blue correspond to the MDC, TOF and EMC of
  the BESIII detector, respectively.}
\label{fig::tracks}
\end{center}
\end{figure}

The transverse view of the BESIII detector in Fig. \ref{fig::tracks}
shows an event with $\Lambda\bar\Lambda$ space-like separation. The
neutral hyperon pair originates from the $\eta_c$ decay at the primary
vertex. The decay-length ratio is $L_1/L_2 = 2.711$ for this event. The $p\bar p$ daughter particles fly along curves with large radii, and the $\pi^\pm$ along curves with smaller radii. The momenta of the hyperons are calculated from those of their daughter particles, and $S= 4.854$ here, satisfying the space-like separation criterion.
%%%%%%%%%%%% figure 1 %%%%%%%%%%

\section{AMPLITUDE ANALYSIS}
An amplitude analysis is used to calculate the probability of
events with $\gamma\Lambda\bar{\Lambda}$ final states for the
selected events, which are divided into three categories. The first
category refers to background events in which the final states are
misidentified as $\gamma\Lambda\bar\Lambda$; the second corresponds to spin-entangelment signal events including the intermediate state $\eta_c$ and the
non-resonant ($NR$) case with quantum numbers $J^P=0^-$, denoted by
$NR(0^-)$; the last category is due to spin-entanglement background
events, i.e., $J/\psi\to\gamma
NR(0^+,1^+,2^+)\to\gamma\Lambda\bar\Lambda $ decays. The amplitudes
corresponding to $NR(0^\pm,1^+,2^+)$ with $J^P$ are denoted by
$M^{J^P}$, which are obtained by multiplying the helicity amplitudes
of all steps of the cascade decay.

An amplitude model is used to isolate the signal events from the $J/\psi$ cascade decay, in which the probability of finding the intermediate state $J^P$ is calculated by evaluating the weight factor of the MC event $i$, which is defined as,
\begin{equation}\label{eq::wts}
W_i(\vec \xi,\vec \omega_i) = {1\over \mathcal N}\overline\sum|M^{J^P}(\vec \xi,\vec \lambda,\vec \omega_i)|^2{N_\text{dt}-N_\text{bg}\over N_\text{MC}},
\end{equation}
where $\overline\sum$ denotes the summation over the helicities of the photon, proton and antiproton, and taking the average over the spin third-component of the J/$\psi$. The variables $N_\text{dt}$, $N_\text{bg}$ and $ N_\text{MC}$ denote the numbers of data, background and MC phase-space events, respectively. $\mathcal N$ is a normalization factor calculated as the amplitude squared average of $N_\text{MC}$ events. The vector $\vec \lambda$ denotes the helicities of the particles involved. The parameter $\vec\xi$ is determined by fitting the amplitudes to the data events with 9-dimensional vectors $\vec\omega_i$.

The mass spectra of $\gamma\Lambda/\gamma\bar{\Lambda}$ and $\Lambda\bar\Lambda$ can be well described by $NR$ in conjunction with the $\eta_c$ signal, and the statistical significance of $\eta_c$ is determined to be more than 5$\sigma$. The criterion to include $NR$ is that its statistical
significance is more than $5\sigma$, and the spin and parity of
these states surviving the event selection criteria are determined to be $J ^ P = 0 ^
\pm, 1 ^ +$, or $2 ^ +$. However, only the $\Lambda\bar\Lambda$ pairs produced by the decays of $\eta_c$ and $NR(0^-)$, totaling 14716 observed events, are in the spin entangled state being studied. The contributions from other non-resonant states with $J^P=(0^+, 1^+, 2^+)$ constitute the entanglement background.(More details could be found in the Appendix)
\section{RESULT}
\label{sec:sc}
\begin{figure}[htbp]
\begin{center}\vspace{1cm}
\begin{overpic}[width=0.6\textwidth]{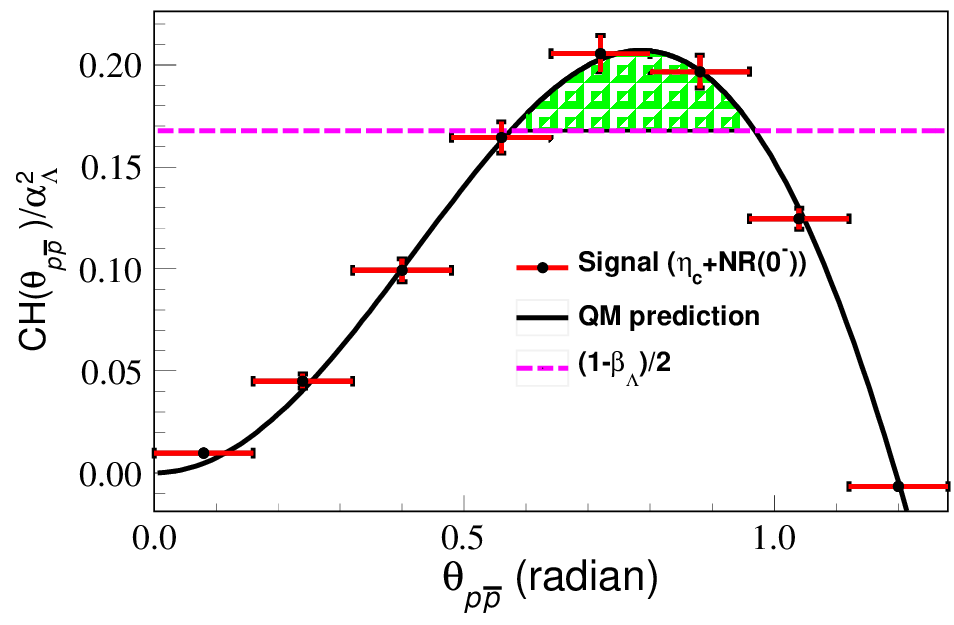}
\end{overpic}
\caption{{\bf| The distribution of $CH (\theta _{p\bar p}) / \alpha_{\Lambda}^2 =[\frac{3\cos\theta_{p\bar{p}}-\cos(3\theta_{p\bar{p}})}{4}-\frac{1}{2}]$.} The points with total error bars are the measurements, the solid line is the QM prediction, and the dashed line is the upper bound of the LHVT prediction. The shaded area above the dashed line indicates the violation of the $CH (\theta _{p\bar p})$ inequality.}
\label{fig::ch}
\end{center}
\end{figure}

To test the realism, the distribution of $CH(\theta_{p\bar{p}})$ for the $\Lambda\bar\Lambda$ spin-entangled
events are measured, where the background and entanglement background are subtracted using simulated events weighted to agree with the amplitude analysis solution. More detail can be found in the Appendix. The distribution of $CH(\theta_{p\bar p})/\alpha_{\Lambda}^2$ for the $\Lambda\bar{\Lambda}$ spin-entanglement events is shown in Fig.\ref{fig::ch}. The points with total error bars, corresponding to signal, are the numbers of events with simulated background and weighted $NR(0^+,1^+,2^+)$ events subtracted in each bin, corrected by the detection efficiency times the value of $CH(\theta_{p\bar p})/\alpha_{\Lambda}^2$. For comparison with the theoretical distribution, the points are further scaled by the ratio of the area under the signal $CH(\theta_{p\bar p})/\alpha_{\Lambda}^2$ distribution divided by that under the theoretical $CH(\theta_{p\bar p})/\alpha_{\Lambda}^2$ distribution. The events in the shaded region above $\frac{(1-\beta_{\Lambda})}{2}$ line are consistent
with the QM prediction~\cite{Qian:2020}, and they are located above the upper bound of the LHVT prediction, indicating that the CH inequality is significantly violated. We obtain $\chi ^ 2 = \sum_ {i
= 1} ^ 4 ( N_i -U_b) ^ 2/\sigma_i^2 = 30.9$ for the two bins in
this interval, where $N_i(\sigma_i)$ denotes the measurement
(total uncertainty) of the $i$th bin, and $U_b$ is the upper boundary of
LHVT~\cite{Qian:2020}. This shows the significance of rejecting the
CH inequality is determined to be 5.2$\sigma$.

We also test the Bell and Clauser–Horne–Shimony–Holt (CHSH) inequalities in two QM dependent schemes, with further details provided in the Appendix. To measure the $p\bar p$ angular distribution, the significance to exclude the Bell inequality region is determined to be $8.9\sigma$. The measurements are consistent with the QM predictions and clearly contradict the predictions of the LHVT. We check the CHSH inequality by calculating the $C_{ij}$ tensor using a TOY MC method based on amplitude analysis. A $\chi^2$ test shows that excluding the LHVT is significant at a level exceeding $10\sigma$.\\
%%%%%%%%%%% figure 2 %%%%%%%%%%%%
Using $(10.087\pm0.044)\times10^{9}$ $J/\psi$ events collected with
the BESIII experiment, a non-local correlation test for the spin
entanglement of a hyperon system in
$J/\psi\to \gamma\eta_c,\eta_c\to\Lambda\bar\Lambda$ decays is carried
out for the first time. The decay lengths of the $\Lambda$ and $\bar\Lambda$
particles, detectable at a macroscopic scale, are used to control the
selection of space-like separation events, which ensures that the
locality loophole is closed. The $CH (\theta_{p\bar
  p})$ angular distributions between the proton and antiproton are in
agreement with the QM predictions within their uncertainties. The Bell and CHSH inequalities are also tested, which however are QM dependent. The three tests have significantly excluded LHVT,
confirming the existence of non-local quantum correlations, with the
significances of $5.2\sigma$, $8.9\sigma$ and larger than $10\sigma$ for the
CH, Bell and CHSH inequalities, respectively.
These results confirm the existence of quantum entanglement and the violation of the Bell inequality in the presence of strong and weak interactions. It tells that the entanglement emerging from these fundamental interactions also exhibits quantum correlation and nonlocality, which deepens our understanding of the physical reality.
\section*{Acknowledgements}
The BESIII Collaboration thanks the staff of BEPCII and the IHEP computing center for their strong support. This work is supported in part by National Key R\&D Program of China under Contracts nos 2020YFA0406300 and 2020YFA0406400; National Natural Science Foundation of China (NSFC) under Contracts nos 12175244, 12275058, 12235008, 12475087, 11875115, 11875262, 11635010, 11735014, 11835012, 11935015, 11935016, 11935018, 11961141012, 12022510, 12025502, 12035009, 12035013, 12061131003, 12192260, 12192261, 12192262, 12192263, 12192264, 12192265, 12221005, 12225509 and 12235017; the Chinese Academy of Sciences (CAS) Large-Scale Scientific Facility Program; the CAS Center for Excellence in Particle Physics (CCEPP); Joint Large-Scale Scientific Facility Funds of the NSFC and CAS under Contract no. U1832207; CAS Key Research Program of Frontier Sciences under Contracts nos. QYZDJ-SSW-SLH003 and QYZDJ-SSW-SLH040; 100 Talents Program of CAS; The Institute of Nuclear and Particle Physics (INPAC) and Shanghai Key Laboratory for Particle Physics and Cosmology; ERC under Contract no. 758462; European Union's Horizon 2020 research and innovation programme under Marie Sklodowska-Curie grant agreement under Contract no. 894790; German Research Foundation DFG under Contracts nos 443159800 and 455635585, Collaborative Research Center CRC 1044, FOR5327, GRK 2149; Istituto Nazionale di Fisica Nucleare, Italy; Ministry of Development of Turkey under Contract no. DPT2006K-120470; National Research Foundation of Korea under Contract no. NRF-2022R1A2C1092335; National Science and Technology fund of Mongolia; National Science Research and Innovation Fund (NSRF) via the Program Management Unit for Human Resources \& Institutional Development, Research and Innovation of Thailand under Contract No. B16F640076; Polish National Science Centre under Contract no. 2019/35/O/ST2/02907; The Swedish Research Council; U. S. Department of Energy under Contract no. DE-FG02-05ER41374.

\vspace{1cm}
\appendix
\renewcommand{\thesection}{\arabic{section}}
\centerline{\bf \LARGE APPENDIX}

\section{Event selection}
\begin{figure}
    \centering
    \includegraphics[width=0.8\textwidth]{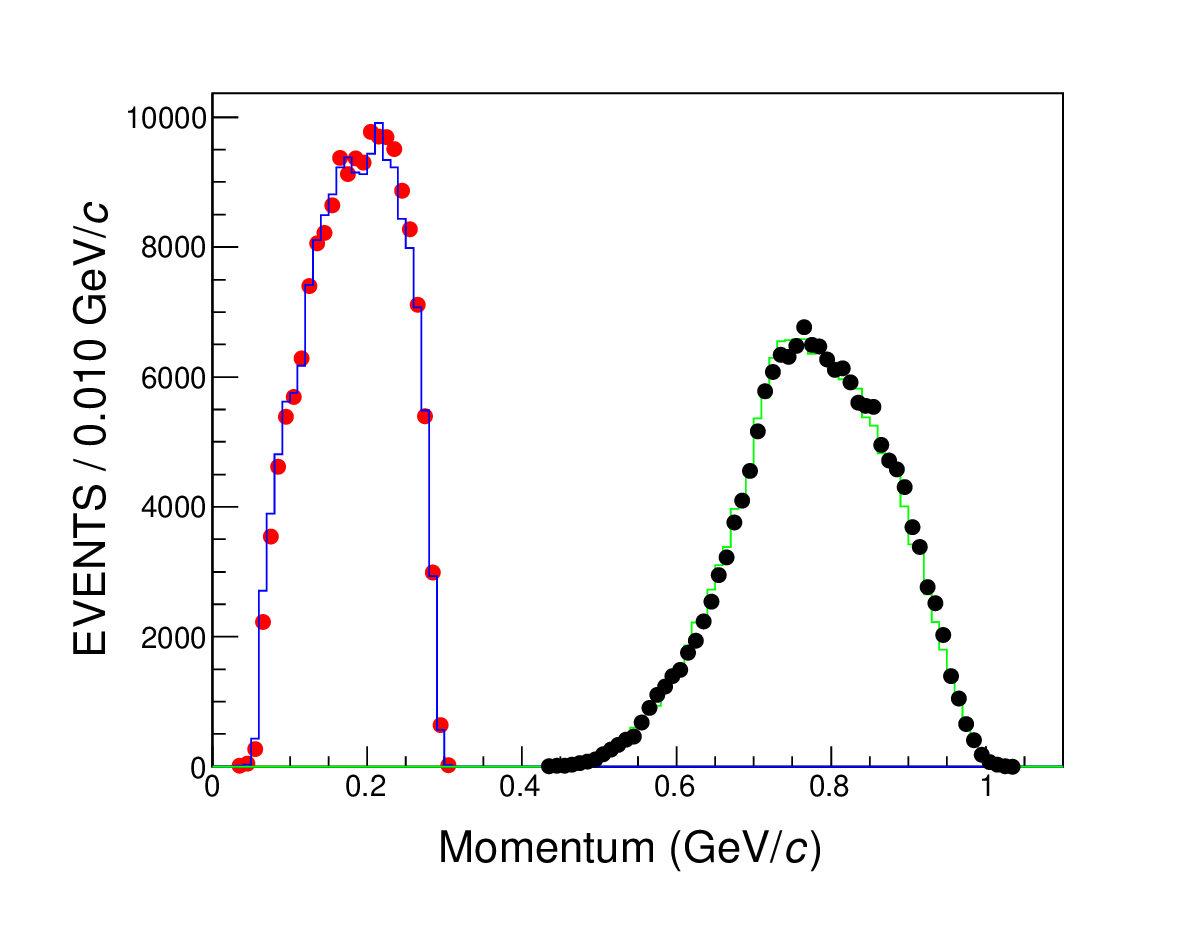}
    \caption{Comparison of the momentum distributions
        of pions and protons for the decay
        $J/\psi\to\gamma\eta_c\to\gamma\Lambda\bar{\Lambda}\to\gamma
        p\pi^-\bar{p}\pi^+$ with signal MC events. The black dots and
      green line denote data and MC events, respectively,
      corresponding to $p$ ($\bar p$), and the red dots and blue line
      denote data and MC events corresponding to $\pi^+$ ($\pi^-$).}
    \label{fig:mom}
\end{figure}

\begin{figure}
\begin{center}
\begin{overpic}[width=0.8\textwidth]{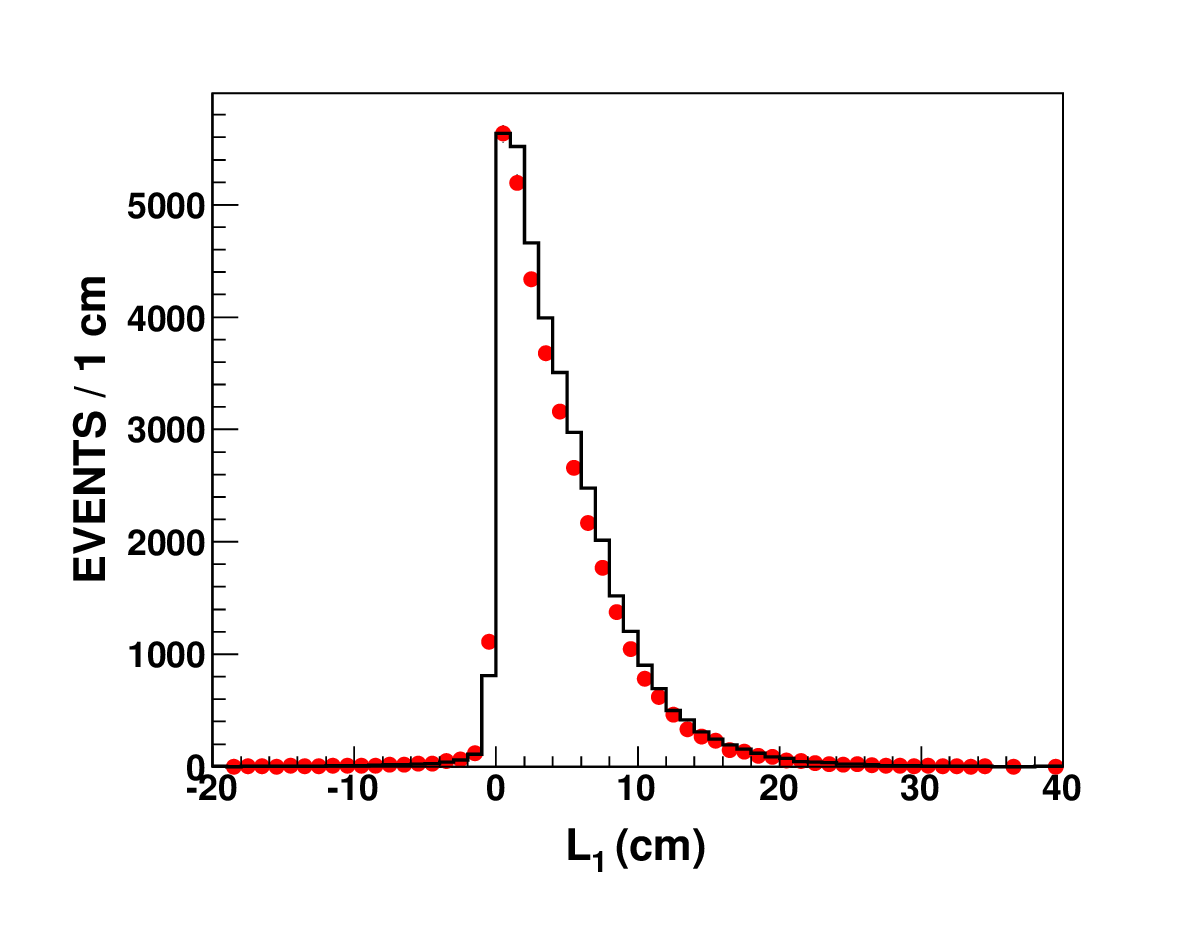}\put(25,60){(a)}
\end{overpic}
\begin{overpic}[width=0.8\textwidth]{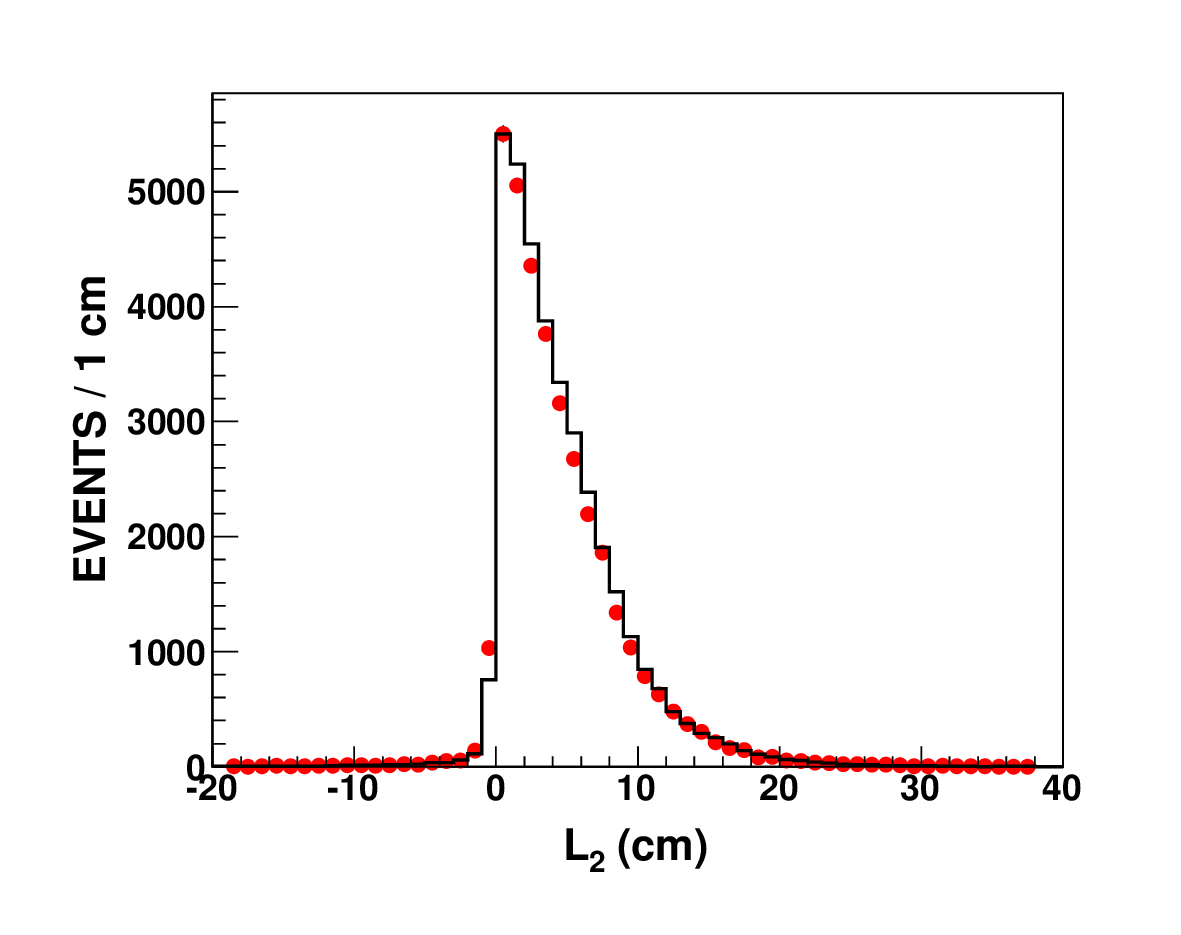}\put(25,60){(b)}
\end{overpic}\\
\caption{The comparison of decay lengths of (a) $\Lambda$ and (b) $\bar \Lambda$. The red dots denote data, and the black histograms denote the signal MC sample.}
\label{fig:declen}
\end{center}
\end{figure}

The final state of a candidate event is required to contain four charged tracks ($p, {\bar p}, \pi^+$ and
$\pi^-$) and at least one good photon ($\gamma$). The charged tracks
reconstructed in the MDC are required to satisfy
$|\cos\theta|\le0.93$, where $\theta$ is defined with respect to the
$z$-axis which is the symmetry axis of the MDC. The momentum
distributions of protons and pions from the signal process are well
separated and do not overlap, as shown in Fig.~\ref{fig:mom}. Therefore, a simple momentum criterion is applied:
if the momentum of the particle is larger than 0.4 GeV/$c$, it will be
identified as a proton, otherwise it will be identified as a pion. Next,
the $\Lambda \to p\pi^-$ and $\bar\Lambda\to \bar p \pi^+$ decays are
reconstructed, requiring for each that two tracks with opposite charges can be
successfully fit to a secondary vertex and requiring the invariant
mass of the hyperon lies in the interval of [1.008,1.124] GeV/$c^2$. The
hyperon decay length distributions of the MC events are in good
agreement with the data (shown in Figs.~\ref{fig:declen}).

Photon candidates are
reconstructed from showers in the EMC within 700 ns from the event
start time. The deposited energy of each shower is required to be
greater than 25 MeV in the barrel region ($|\cos\theta|\le0.8$) or 50
MeV in the end cap region ($0.86\le|\cos\theta|\le0.92$), and the
minimum opening angle between the shower and the pion or nucleon
(antinucleon) is required to be greater than $20^\circ(30^\circ)$. The
radiative photon is selected through a four constraint (4C) kinematic
fit requiring energy and momentum conservation in the decay
$J/\psi\to\gamma\Lambda\bar{\Lambda}$, and events with $\chi^2_{\rm
  4C}(\gamma\Lambda\bar\Lambda)<\chi^2_{\rm
  4C}(\gamma\gamma\Lambda\bar\Lambda)$ and $\chi_{\rm
  4C}^2(\gamma\Lambda\bar\Lambda)< 30$ are retained for further
analysis, where $\chi_{\rm 4C}^2$ is the goodness of fit of the
kinematic fit. In order to remove background events containing
$J/\psi\to \Sigma^0\bar\Sigma^0,\Lambda\bar\Sigma^0+c.c.$ decays, the
invariant mass of $\gamma\Lambda/\gamma\bar\Lambda$ is required to
satisfy $|M_{\gamma\Lambda/\gamma\bar\Lambda}-M_{\Sigma^0}|>0.009$
GeV$/c^2$, where $M_{\Sigma^0}$ is the known mass of $\Sigma^0$
\cite{ParticleDataGroup:2022pth}.

\section{Amplitude analysis using maximum likelihood fit}
\begin{figure}
    \centering
    \includegraphics[width=0.8\textwidth]{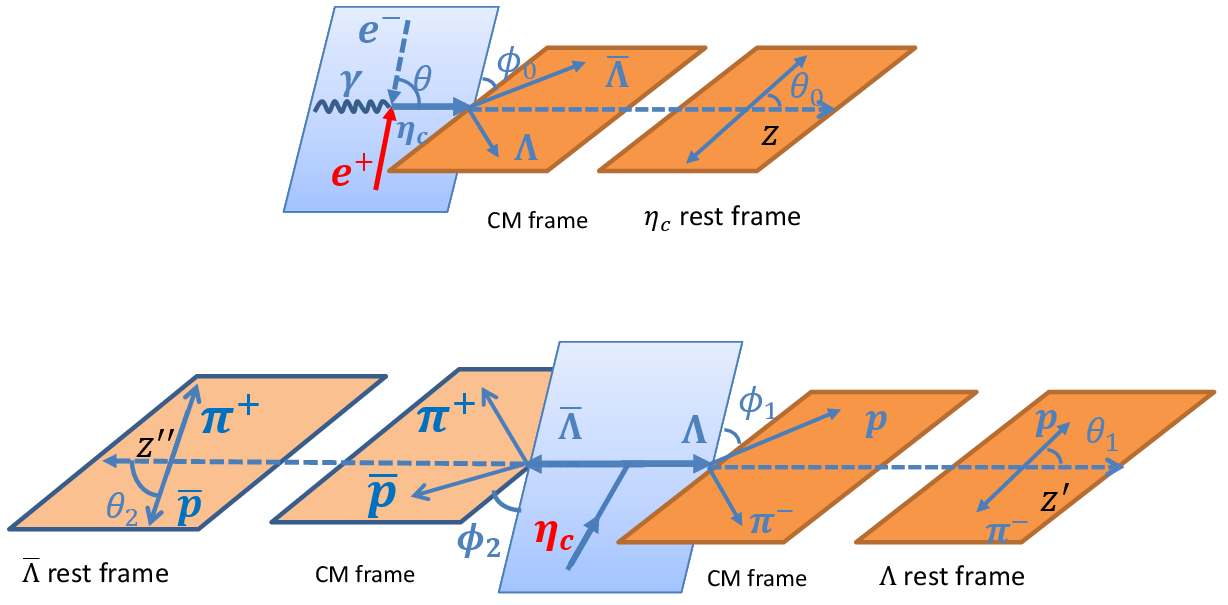}
    \caption{Illustration of helicity angles defined in each step of a decay chain. Helicity angles in $J/\psi\to\gamma\eta_c, \eta_c\to\Lambda \bar\Lambda$  decays, and $\eta_c\to\Lambda \bar\Lambda\to p \bar p \pi^+\pi^-$ decays}
    \label{fig:sysdef}
\end{figure}

The amplitude of the $\eta_c$ or non-resonant ($NR$) decays depends on a
9-dimensional vector $\vec\omega = (M_{\Lambda\bar\Lambda},
\theta_\gamma, \phi_\gamma, \theta_\Lambda, \phi_\Lambda,\theta_p,
\phi_p, \theta_{\bar{p}}, \phi_{\bar{p}})$, where
$M_{\Lambda\bar\Lambda}$ is the invariant mass of the
$\Lambda\bar\Lambda$ system and $\theta_i$ and
$\phi_i(i=\gamma,\Lambda,p,\bar p)$ denote, respectively, the polar and
azimuthal angles of particle $i$ in their helicity coordinate systems,
which are illustrated in Fig. \ref{fig:sysdef}. For the
decay $A\to B+C$, the polar angle $\theta$ is defined as the
angle between the momentum vectors $\vec p_A$ and $\vec p_B$, which
are defined in the rest frame of the mother particle. The
azimuthal angle $\phi$ is defined as the angle between the
production and decay planes of particle $A$.

For the cascade decays
$J/\psi(m)\to\gamma(\lambda_\gamma)R(\lambda_0),~R\to\Lambda(\lambda_1)\bar\Lambda(\lambda_2),~\Lambda\to
p(\lambda_3)\pi^-$ and $\bar\Lambda\to \bar p(\lambda_4)\pi^+$, where
$m$ and $\lambda_i$ denote the third component of the $J/\psi$ spin
and the helicity of the particle $i$, respectively, the amplitudes are
expressed as
\begin{eqnarray}
M^{J^P}(\vec \xi,\vec \lambda,\vec \omega)&=& \sum_{\lambda_0,\lambda_1,\lambda_2} D^{1*}_{m,\lambda_\gamma-\lambda_0}(\phi_\gamma,\theta_\gamma,0)
D^{J*}_{\lambda_0,\lambda_1-\lambda_2}(\phi_\Lambda,\theta_\Lambda,0)
D^{1/2*}_{\lambda_1,\lambda_3}(\phi_p,\theta_p,0)
D^{1/2*}_{\lambda_2,\lambda_4}(\phi_{\bar p},\theta_{\bar p},0)\nonumber\\
&\times& BW(M_{\Lambda\bar\Lambda},M_0,\Gamma)H^{J/\psi}_{\lambda_\gamma,\lambda_0}(\vec \xi)H^R_{\lambda_1,\lambda_2}(\vec\xi)H^\Lambda_{\lambda_3,0}(\vec \xi)H^{\bar\Lambda}_{\lambda_4,0}(\vec \xi),
\end{eqnarray}
where $BW (M _{\Lambda\bar\Lambda}, M_0, \Gamma)$ is the relativistic
Breit-Wigner (BW) function describing the $\eta_c$ resonance with a mass of
$ M_0$ and a width of $\Gamma$. For non-resonant transitions, this
$BW$ factor is set to 1. $D_{n, h}^J (\phi, \theta, 0)$ are
elements of the Wigner-$D$ matrix, where $J$ is the spin of $R$ and $n$
and $h$ correspond to helicities. $ H_{\lambda_B, \lambda_C}^A$
denotes the helicity amplitude of the decay $A \to B (\lambda_B) C
(\lambda_C)$, which is expanded into partial waves in terms of
the orbital angular momentum $L$ and the total spin $S$ of the decay,
and combined linearly with the $L$-$S$ coupling parameter $\vec \xi$
\cite{chung2}. For $H_{\lambda_\gamma, \lambda_0}^{J/\psi}$ and
$H_{\lambda _ 1, \lambda_2}^ R$, the number of partial waves is
restricted by parity conservation. For the $\Lambda (\bar\Lambda) \to
p\pi^- (\bar{p}\pi^+)$ weak decays, their amplitudes,
$H_{\lambda_3,0}^\Lambda$ and $H_{\lambda_4,0}^{\bar\Lambda}$, are expanded
in terms of $S$- and $P$-wave amplitudes. Because the two decays
approximately conserve the CP quantum numbers
\cite{BESIII:2018cnd}, the sign of the $S$-waves stays the same,
while the $P$-waves change sign in the charge conjugated
decays. The $L$-$S$ coupling constants involved in all partial wave
amplitudes are set as parameters to be determined by fitting the data.

The probability density of event $i$ is obtained by coherently adding
the amplitudes of all intermediate states, and taking the modulo squared:
\begin{equation}
\sigma_i (\vec\xi,\vec\omega)=\sum_{m,\lambda_\gamma,\lambda_3,\lambda_4}|\sum_{J^P}M^{J^P}(\vec \xi,\vec \lambda,\vec \omega)|^2.
\end{equation}
Here $J^P$ is summed over all resonant and non-resonant states.

We determine the coupling parameters, $\vec\xi$, from a maximum
likelihood fit to data. The likelihood function of an ensemble
with $N$ events is defined as
\begin{equation}
\mathcal{L}(\vec \xi,\vec\omega) = \prod_{i=1}^N \mathcal{P}_i(\vec \xi,\vec\omega)=\prod_{i=1}^N{\sigma_i(\vec \xi,\vec\omega) \over \mathcal{N}(\vec \xi)},
\end{equation}
where $\mathcal{N}=\int\sigma_i(\vec
\xi,\vec\omega)d\vec\omega$ is the
normalization factor, which accounts for the detection and
reconstruction efficiency and is approximated as the MC integral i.e.,
the average value of the integrand is estimated with a sufficiently
large number of MC events. Here the MC events are generated with a
phase-space model, subjected to detector simulation, and required to
survive the event selection criteria. The minimum of the objective
function
\begin{equation}
S=-[\ln\mathcal{L}(\vec \xi,\vec\omega_{\rm dt})-\ln\mathcal{L}(\vec \xi,\vec\omega_{\rm bg})]
\end{equation}
corresponds to the maximum of the likelihood function
$\mathcal{L}$. To obtain the coupling parameters $\vec\xi$ in the
amplitude analysis, $S$ is minimized with {\sc
  minuit2}~\cite{James:1975dr}, and the contribution from the
background events obtained from the exclusive MC samples, $\ln\mathcal{L}(\vec \xi,\vec\omega_{\rm bg})$, is
subtracted from the objective function of the data
$\ln\mathcal{L}(\vec \xi,\vec\omega_{\rm dt})$.  The dominant background mainly consists of $J/\psi\to\bar{\Lambda}\Sigma^{0}$ + $c.c.$, $J/\psi\to\bar{\Lambda}(1520)\Lambda\to\gamma\bar{\Lambda}\Lambda$ + $c.c.$, and $J/\psi\to\bar{\Sigma}^{0}(\gamma\bar{\Lambda})\Sigma^{0}(\gamma\Lambda)$, with their contributions estimated using MC samples.
The efficiency of the data obtained in 2012 is higher than that of the other runs by 14\%, since the MDC magnetic field setting was lower than that in other years. Consequently, the dataset of 10 billion data events is divided into two sub-samples and then fitted simultaneously to determine the
parameters. Approximately 11\% of the data was obtained with the lower magnetic field.

\begin{figure}
    \centering\vspace{1.0cm}
    \includegraphics[width=0.8\textwidth]{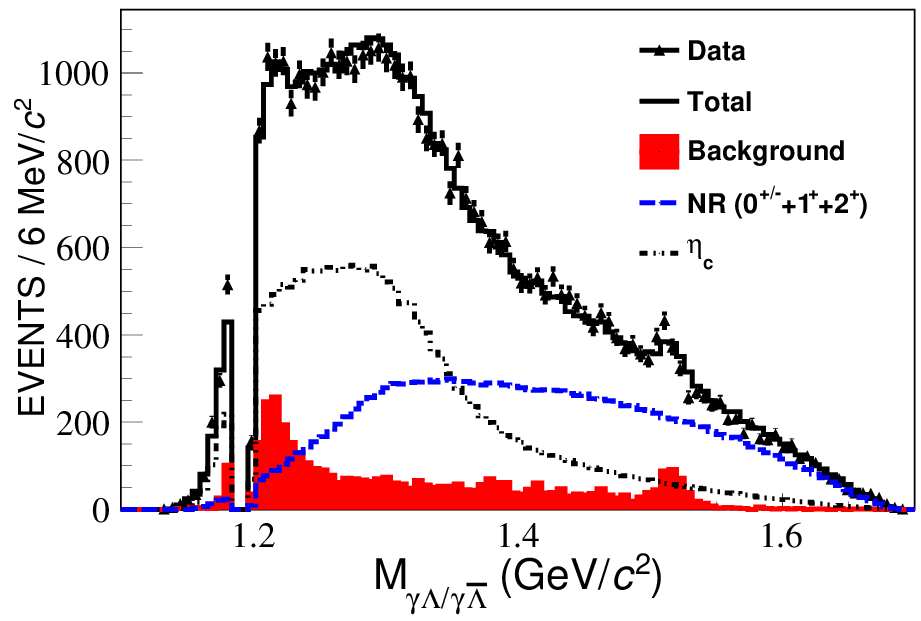}
    \caption{The invariant mass spectrum $M_{\gamma\Lambda /\gamma\bar\Lambda}$ (one event is filled twice by $M_{\gamma\Lambda}$ and $M_{\gamma\bar\Lambda})$. The triangles with error bars are data, the black histogram is the total fit result, the red shaded part is the background, the blue dash line is the $NR$ state, and the black dash-dotted line is the signal $\eta_c$.}
    \label{fig:mgL}
\end{figure}

\begin{figure}
    \centering
    \includegraphics[width=0.8\textwidth]{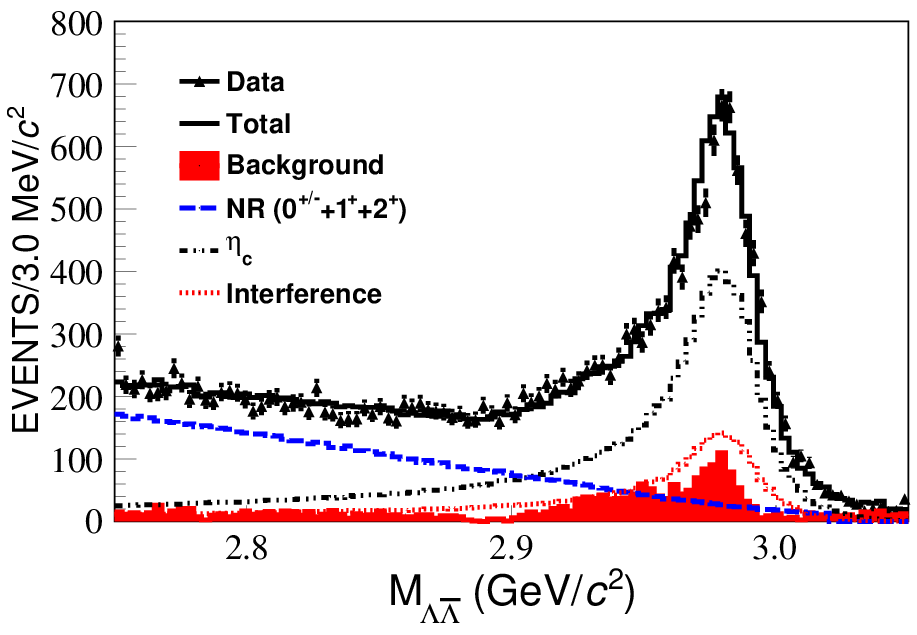}
    \caption{The invariant mass spectrum $M_{\Lambda\bar\Lambda}$. Along with the items described in Fig. \ref{fig:mgL}, the interference of the $\eta_c$ with $NR(0^\pm, 1^+, 2^+)$ is shown.}
    \label{fig:final}
\end{figure}

\begin{figure}
    \centering
    \includegraphics[width=0.8\textwidth]{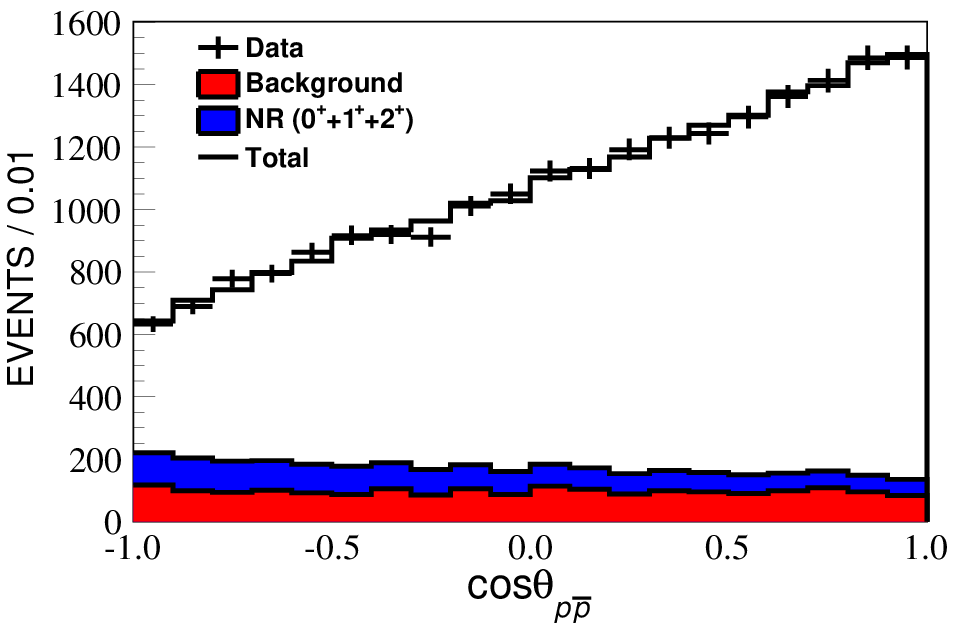}
    \caption{The angular distribution $\cos\theta_{p\bar p}$. The triangles with error bars  are data, the histogram is the total fit result, and the shaded areas are the background and non-resonant states.}
    \label{fig:angEPR}
\end{figure}

\begin{figure}
    \centering
    \vspace{1.0cm}
    \includegraphics[width=0.6\textwidth]{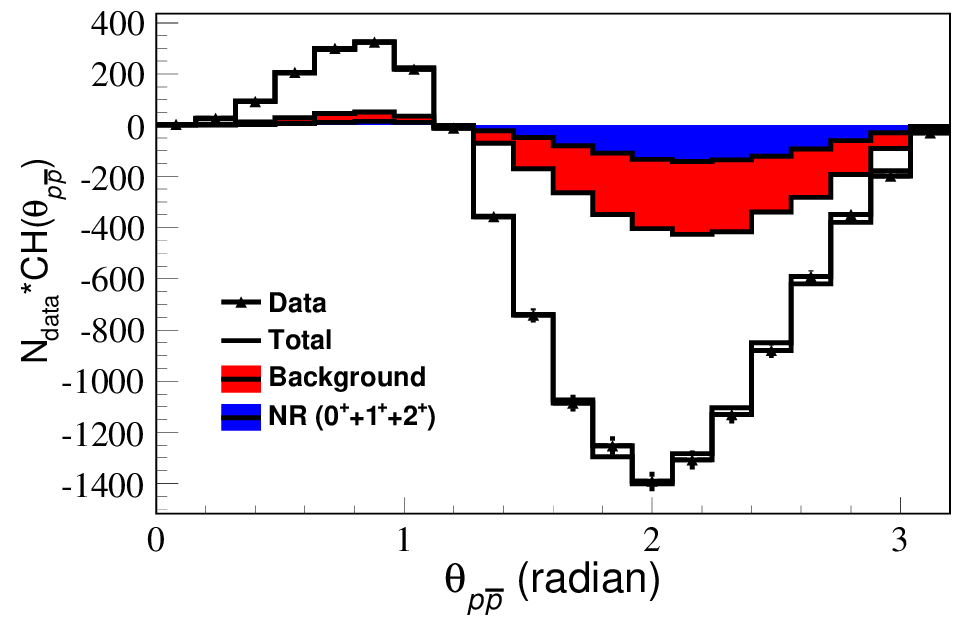}
    \caption{The distribution of $CH(\theta_{p\bar p})=\alpha^2_{\Lambda}[\frac{3 \cos\theta_{p\bar{p}}-\cos(3\theta_{p\bar{p}})}{4}-\frac{1}{2}]$. The legend is the same as Fig. \ref{fig:angEPR}.}
    \label{fig:angCH}
\end{figure}

To have the signal and $NR$ samples agree with the solution obtained from the amplitude model, phase space MC events are weighted by $\mathcal{P}_i(\vec \xi,\vec\omega)$, with parameters obtained from the maximum likelihood fit. Background is made up of inclusive MC events. The numbers of simulated events used is obtained from the maximum likelihood fit.The invariant mass spectra of $\gamma\Lambda$($\gamma\bar\Lambda$) and ${\Lambda\bar\Lambda}$ are displayed in Figs.~\ref{fig:mgL} and \ref{fig:final}. The signal $\eta_c$ and $NR(0^\pm,1^+,2^+)$ components are parameterized by weighted phase space.\\
\vspace{2.0cm}
Figure~\ref{fig:angEPR} shows the $\cos\theta_{p\bar p}$ distribution. The total histogram is the sum of weighted simulated samples of signal entangled events ($\eta_c$ and $NR(0^-)$), $NR(0^+, 1^+, 2^+)$, and background. Figure~\ref{fig:angCH} shows the $CH(\theta_{p\bar{p}})$ distribution multiplied by the number of data events. The entries are the number of $\eta_c$ signal events in each bin times the value of $CH$ for that bin. The number of signal events is given by the data minus $NR(0^+, 1^+, 2^+)$ and background.
The yields of $J^P$ components are determined according to the MC event weights, namely, the ratio of cross section for the $J^P$ component over the total cross section. Supplementary Table~\ref{tab::yields} shows the number of yields in data for the $NR(0^+,0^-,1^+,2^+)$ and $\eta_c$ components.
\begin{table}
\centering
\caption{The yields of each component determined in the amplitude fit.\label{tab::yields}}
\begin{tabular}{cccccc}
\hline\hline
Components & $NR(0^+)$ & $NR(1^+)$ & $NR(2^+)$ & $NR(0^-)$ & $\eta_c$\\
\hline
Yields & $892.4\pm 230.3$ & $116.2\pm 52.8$ & $478.5\pm 155.6$ & $6139.7\pm 353.2$ & $8576.6\pm 154.6$ \\
\hline\hline
\end{tabular}
\end{table}
\subsection{Test of Bell inequality}
\begin{figure}
\begin{center}\vspace{2.0cm}
\begin{overpic}[width=0.8\textwidth]{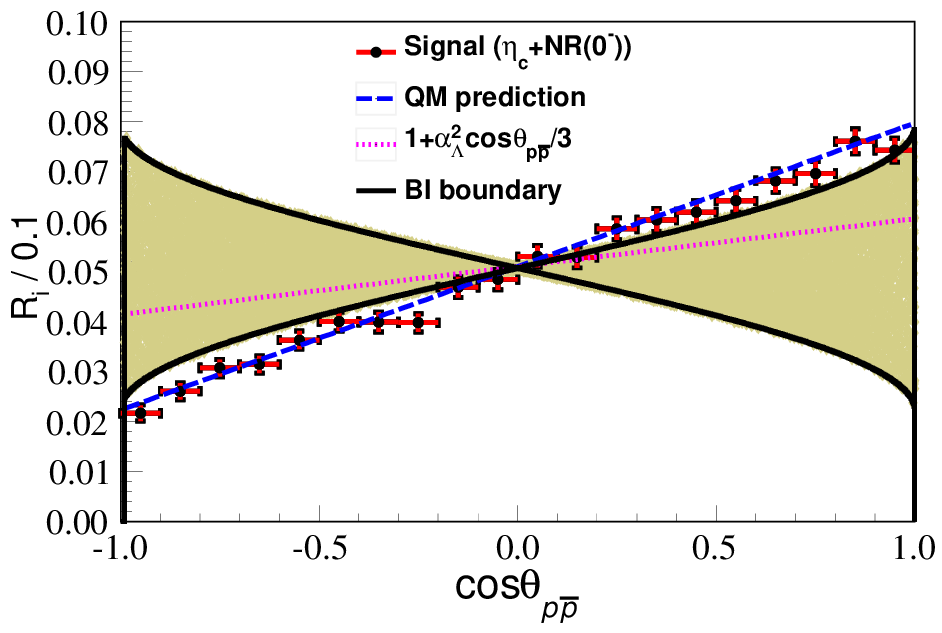}
\end{overpic}
\caption{Normalized signal event distribution ($R_i = N_i/N_{tot}$) of $\cos\theta_{p\bar p}$. Here $N_i(N_{tot})$ is the number of events in $i$-th bin (total signal events). The points with total error bars are the signal, while the dashed line is the QM prediction, with a slope of $\alpha_\Lambda ^ 2$. The dotted line represents the QM prediction with $\Lambda$ polarization determined by a hidden process. The filled region indicates the domain that satisfies the Bell inequality (denoted by BI in this figure) in Eq.(3).}
\label{fig::epr}
\end{center}
\end{figure}

The angular distribution of $p\bar p$ is shown in Fig. \ref{fig::epr}. The points with total error bars, corresponding to signal, are obtained by subtracting the background and the weighted spin-entanglement background events in the amplitude model from the selected events, correcting by the efficiency, then normalizing by the total number of signal events. The dashed line shows the QM prediction with a slope of $\alpha_\Lambda^2$, we take $\alpha_\Lambda = 0.750\pm0.009\pm0.004$~\cite{BESIII:2018cnd}. The dotted line represents the QM prediction with $\Lambda$ polarization determined by a hidden process, and its slope is $\alpha^2_\Lambda/3$. The shaded region represents the region that satisfies the Bell inequality of Eq. (3) in the main text. However, the measured angular distribution is outside this region, indicating a significant violation of the Bell inequality, while being consistent with the QM prediction for $\Lambda\bar\Lambda$ spin-entanglement events. To assess the significance of distinguishing the QM prediction from the Bell inequality, we fit the measured distribution using a linear combination of both the nearby boundary of the Bell inequality and the QM prediction (refer to Fig. \ref{fig::epr}), with two parameters. The change in the $\chi^2$ of the binned fit, with and without the QM prediction, is 69.2, corresponding to a change of one degree of freedom with consideration of systematic uncertainties. Therefore, the significance to exclude the Bell inequality region is $8.9\sigma$.\\
%\vspace{0.3cm}

\subsection{Test of Clauser-Horne-Shimony-Holt inequality}
Another type of Bell inequality was proposed based on the CHSH inequality. Utilizing the bilinear expression of the CHSH inequality in momentum space \cite{Chen:2013epa}, one has
\begin{equation} |\langle n_an_b'\rangle +
\langle n_an_d'\rangle +\langle n_cn_b'\rangle -\langle
n_cn_d'\rangle | \le {2\alpha_\Lambda^2\over 9},
\end{equation}
Here, ${\bf n}={\bf p/|p|}$ and ${\bf n'}={\bf p'/|p'|}$, where ${\bf
  p}$ and ${ \bf p'}$ represent the momenta of the proton and
anti-proton in the rest frames of $\Lambda$ and $\bar \Lambda$,
respectively. The guide directions are denoted by $a$, $b$, $c$, and
$d$, and momentum projections are defined as $\langle n_an_b'\rangle = \langle {\bf n\cdot a}~{\bf n'\cdot d} \rangle $, for example. In the case of maximal violation of the Bell inequality, the momentum correlation can be related to the decay amplitude by defining:
\begin{eqnarray}\label{eq::cij}
C_{ij}\equiv\langle n_in'_j\rangle {9\over 2\alpha^2_\Lambda}
 = {9\over 2\alpha^2_\Lambda|{\bf p}|^2}{\int d\Omega_\Lambda d\Omega_p d\Omega_{\bar p}p_ip_j' |\mathcal{M}_0|^2 \over
 \int  d\Omega_\Lambda d\Omega_p d\Omega_{\bar p} |\mathcal{M}_0|^2
 }, ~~~~ (i,j = 1, 2,3)
\end{eqnarray}
where the indices $i, j$ label the $x, y, z$ components of a vector
in Cartesian coordinates; $\mathcal{M}_0$ represents the amplitude
for the sequential decay $\eta_c\to\Lambda\bar\Lambda,~\Lambda\to
p\pi^-$ and $\bar\Lambda\to \bar p\pi^ +$. While $C_{ij}$ depends on the QM measurement $\alpha^2_\Lambda$, it has been demonstrated
that the correlation tensor can produce distinct maximum values of
$C_{ij}$ for predictions based on QM and local realism (LR), such as:
\begin{eqnarray}\label{eq::qmax}
Q_{max}=2\sqrt{C_{11}^2+C_{33}^2}=\left\{
\begin{array}{cc}
1& (\text{LR}) \\
\sqrt{2}&(\text{QM})
\end{array}
\right..
\end{eqnarray}
The LR predicts a maximum
value of $ Q_{max}=1$, whereas QM can violate
local realism with a maximum value of $\sqrt{2}$.

\begin{figure}
\begin{center} \vspace{2.0cm}
\includegraphics[width=0.8\textwidth]{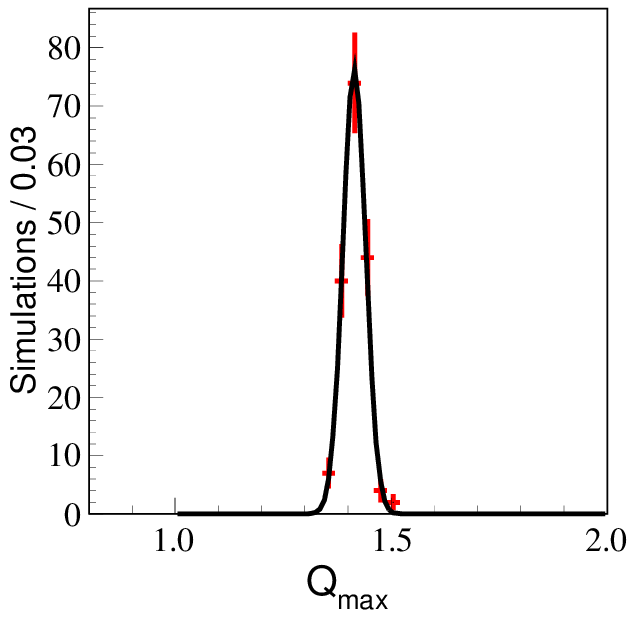}
\caption{The distribution of $Q_{max}$ is
plotted as a function of the number of TOY MC simulations. The dots with error bars represent the TOY MC results, while the line represents the fitted results using a Gaussian distribution.}
\label{fig::qmax}
\end{center}
\end{figure}

We tested the CHSH inequality by calculating the $C_{ij}$ tensor (see Eq.~\ref{eq::cij}) using
the signal of TOY MC events. These events were generated based on the amplitude model with fixed parameters obtained from the amplitude analysis. The statistics of each TOY MC
experiment were fixed to those obtained in the amplitude analysis. We
then calculated the distribution of $Q_{max} $ according to Eq.~\ref{eq::qmax} as a function of the
number of TOY MC experiments, as shown in Fig.~\ref{fig::qmax}. The
distribution was fitted with a Gaussian function, which provided a
good description of the $Q_{max}$ distribution. The mean ($\mu$) and
standard deviation ($\sigma$) of the Gaussian distribution were
determined to be $\mu=1.416\pm0.0 02$ and $\sigma=(2.771\pm 0.001)\times 10^{-2}$,
respectively. The $p$-value is calculated with $Q_{max}<1$. Then the significance of vetoing LHVT was estimated to be larger than $10\sigma$.

%=========================================
%=========================================
\section{Systematic uncertainties}
The systematic uncertainties considered here include: tracking
efficiency, photon detection efficiency, space-like separation
criteria, kinematic fit, background estimation, and the mass and width
of $\eta_c$. To account for potential correlations among the systematic sources, all systematic sources, with the exception of the space-like separation, are collectively considered in an alternative fit, rather than being treated separately. Initially, we adjust the MC sample to accommodate the kinematic fit, followed by corrections for tracking efficiency, photon efficiency, and $\Lambda/\bar\Lambda$ reconstruction, achieved by multiplying their correction factors to the predicted amplitude squared. The uncertainties in the mass and width of $\eta_c$ are factored in when calculating its Breit-Wigner amplitude, by randomly smearing its mass and width. During the subtraction of the background event contribution from the log-likelihood, the weighted factor is smeared in accordance with the statistical uncertainty of the background. The difference in the $\cos\theta_{p\bar{p}}$ distribution between the nominal and alternative fit is considered as the systematic uncertainties. The systematic uncertainties for the $\cos\theta_{p\bar p} $ distribution are negligible. However, the systematic uncertainty of the CH inequality distribution is primarily influenced by the space-like separation requirement. The systematic uncertainties are listed as below.

\begin{itemize}

\item {\bf Tracking efficiency.} The tracking efficiencies of
  $p,\bar p$ and $\pi^\pm$ are obtained by studying the control sample
  $J/\psi\to\Lambda\bar\Lambda\to p\bar p\pi^+\pi^-$. The event
  selection criteria of the control sample are the same as those
  applied in the data analysis. For events with a missing charged track, three charged tracks are identified using PID, and two of the tracks with opposite charge are required to be successfully fitted to a secondary $\Lambda$ vertex. The tracking efficiency of
  reconstructing a charged track $i$ in data and MC events is defined
  as: $\varepsilon(i)={N_{\mathrm{trk}=4}\over
    N_{\mathrm{trk}\ge3}}$. Here $N_{\mathrm{trk=4}}$ is the number of
  events for which all four charged tracks are reconstructed, and
  $N_{\mathrm{trk}\ge3}$ is the number of events, in which all the other
  three charged tracks, in addition to the track $i$ to be studied,
  are reconstructed. The ratio of the efficiency between data and MC
  events is $R_{\mathrm{trk}} (i)={\varepsilon_{\rm data} (i)\over
    \varepsilon_{\rm MC} (i)}$. By comparing the control samples, the
  two-dimensional distributions of $R_{\mathrm{trk}} (p_t,\cos\theta )$
  for $p,\bar p$ and $\pi^\pm$ are determined, where $p_t$ is
  the transverse momentum. In an
  alternative fit, the normalization factor $\mathcal N(\vec\xi)$ is
  multiplied by $R_{\mathrm{trk}} (i)$ for each charged track per
  event. The difference between the alternative and nominal fits is
  taken as the systematic uncertainty.

\item {\bf Photon detection efficiency.} The photon detection efficiency is determined by studying the control sample $J/\psi\to\pi^+\pi^-\pi^0$, as done in
\cite{BESIII:2015rug}. The photon selection criteria for the control
sample are the same as those used in this analysis. The data selection
process requires that each event must contain two oppositely charged
tracks and at least one good photon. In cases where an event contains
multiple good photons, every combination of $\gamma\pi^+\pi^-$ is
reconstructed. Subsequently, energy conservation is then utilized to
pinpoint the region in the EMC where the second photon from $\pi^0 \to
\gamma \gamma$ is expected to be located.  The detection efficiency of
photons is defined as $\varepsilon_{\gamma}={N_{\gamma\gamma}\over
  N_\gamma}$, where $N_{\gamma\gamma}$ and $N_\gamma$ represent the
numbers of events in which two photons and one photon appear, respectively. The
difference in photon detection efficiencies between data and MC events
is determined to be $0.5\%$ and $1.5\%$, corresponding to the barrel
and end cap regions of the EMC, respectively. In an alternative fit,
the normalization factor $\mathcal N(\vec \xi)$ is calculated by
multiplying a Gaussian dispersion factor $\text{Gaus}(1,\delta)$ for
each MC event, where $\delta=0.015$ or $0.005$ depending on whether the
photon is in the barrel or end cap. The difference between the
alternative and nominal fits is taken as the systematic uncertainty.

\item {\bf Space-like separation criteria.} The uncertainty in
detection efficiency due to the space-like separation criteria is
estimated by the control sample $J/\psi\to\Lambda\bar\Lambda$. The
event selection criteria of the control sample are the same as those
applied in the data analysis. Out of $N$ events, $N'$ events satisfy
the requirements of the space-like separation criteria, and the
selection efficiency is defined as $\epsilon = N'/N$. The difference
in selection efficiencies between data and MC events is 0.2\%, which
is negligible.

However, due to the resolution effect of the detector on the measured decay
length of $\Lambda(\bar\Lambda)$ particles, the space-like separation
criteria $L_1/L_2$ is not uniquely determined. The deviation between
the measured value and the true value is determined using MC
simulation, where $\Delta(L_1/L_2)$ follows a Gaussian distribution
with a mean of 0 and a standard deviation of 0.062 cm. To refine our
analysis, we narrowed the window of the $L_1/L_2$ space-like separation
by twice the standard deviation at both ends, and re-selected the
data. The difference between the $p\bar p$ angular distribution and
the CH inequality distribution, obtained by fitting this refined
sample, is considered as the systematic uncertainties in comparison to
the nominal fitting.

\item {\bf Kinematic fit.} The systematic uncertainty of the 4C
  kinematic fit is obtained by applying track parameter
  corrections. The reason for the difference between the kinematic fit
  of the data and MC events is that the pull distributions of the
  helix parameters of charged tracks are not consistent. The
  momentum of the MC event is modified using the
  pull distribution of the helix parameters. In an alternative fit, we
  calculate the normalization factor $\mathcal N(\vec \xi)$ with the
  modified MC events. The difference between the nominal
  and alternative fits is taken as the systematic uncertainty.

\item {\bf Background estimation.} The number of background events estimated by MC samples for the two simultaneously fitted samples, collected in 2012 and other years, are $1,029\pm33$ and $3,290\pm63$. Their statistical fluctuations are used to estimate the uncertainty due to the background. The likelihood function of background events is calculated by multiplying a Gaussian weighting factor $\text{Gaus}(1,\delta n/n)$, where $\delta n/n$ is the relative error of the background. Thus $\ln\mathcal{L}(\vec\xi,\vec\omega_{\rm bg})=\text{Gaus}(1,\delta n/n)\sum_{i=1}^{N_{\rm bg}}\ln\mathcal{P}_i(\vec\xi,\vec\omega_{\rm bg})$ is used in an alternative fit.
\item {\bf $\eta_c$ mass and width.} In the nominal fit, the $\eta_c$ mass and width are fixed to the world averages~\cite{ParticleDataGroup:2022pth}, i.e. $M_0=(2983.9\pm0.4)\text{ MeV}/c^{2}$ and $\Gamma=(32.0\pm0.7)$ MeV. Their relative uncertainties are $\delta M_0/M_0=1.7\times10^{-4}$ and $\delta\Gamma/\Gamma=2.2\%$. The uncertainties are estimated by smearing the $\eta_c$ mass and width in the calculation of the likelihood function by a Gaussian distributions, $M_{\eta_c}\sim \text{Gaus}(M_0,\delta M_0/M_0)$ and $\Gamma_{\eta_c }\sim\text{Gaus}(\Gamma,\delta\Gamma/\Gamma)$.

\end{itemize}

\end{document}